\documentclass[aps,pra,twocolumn,groupedaddress,showpacs,floatfix]{revtex4}

\usepackage{graphicx}    
 \usepackage{amssymb}

\begin{document}
\title[IR Xe excimer emission in dense Xe gas]{Dielectric screening--, multiple scattering--, and quantum indistinguishability effects\\ in the IR emission spectrum of Xe$_{2}$ excimers in dense 
 gas}
\author{A. F. Borghesani}
\affiliation{Department of Physics, CNISM--Unit, University of Padua, and \\ Istituto Nazionale di Fisica Nucleare, Sezione di Padova, Padua, Italy}
\author{G. Carugno}
\affiliation{Istituto Nazionale di Fisica Nucleare, Sezione di Padova, Padua, Italy}
\date{\today}
\begin{abstract}
We report new measurements of the infrared emission spectrum of Xe$_{2}$ excimers produced by electron impact in dense gas at room temperature. These measurements extend the density range previously explored by a factor of two. We have now reached a pressure $P\approx 3.3\,$MPa, corresponding to a gas density $N\approx 8\cdot 10^{26}\,$m$^{-3}$ in both pure Xe gas and in a Ar--Xe mixture. The analysis of the changes of the spectrum induced by the interaction with the gaseous environment shows how cooperative-- and many-body effects develop with increasing density. We 
also show that quantum effects related to the indistinguishability of identical particles are observed even at such high temperature.
\end{abstract}

\maketitle

\section{Introduction}\label{sec:intro}
The luminescence of excited rare gases (RG) is a very well known phenomenon that is mainly studied in the 
 vacuum ultraviolet (VUV) range because of many 
 applications including 
 the production of coherent and incoherent VUV light sources \cite{rhodes1979,ledru2006} and the detection of ionizing radiation \cite{knoll}. 

The luminescence of RG excimers is also interesting because of the knowledge that can be obtained about the structure and the formation dynamics of these diatomic molecules. These pieces of information are relevant in several fields of physics, including molecular--, plasma--, environmental--, and astrophysics.

Xenon is the most accurately studied gas because of its ability to efficiently convert the excitation energy in intense VUV radiation \cite{lorents1976}. Upon excitation accomplished by using several techniques, including high-energy particles \cite{arai1969,koe1974,keto1979,leichner1976,millet1978,davide1}, electrical discharges \cite{colli1954,wieme1974,sadeghi1977}, multiphotonic absorption by means of a laser \cite{broadmann1976, broadmann1977,dehmer1986,gornik1981,gornik1982,salamero1984,moutard1986,alekseev1999,ledru2006}, and synchrotron radiation \cite{broadmann1976,dutuit1978,wenck1979,bonifield1980}, three-body collisions of excited and ground-state atoms lead to the formation of excited dimers whose decay toward the dissociative ground state yields the 1st and 2nd VUV continua at approximately 150 and 170 nm.

The 1st continuum is produced at low gas pressure, $P<20\,$kPa and is originated by radiative transitions from the vibrationally excited $(0_{u}^{+})_{v^{\prime}\gg 0}$ molecular state, correlated with the resonant 6s$\,(^{3}\!P_{1})$ atomic state, toward the dissociative $0_{g}^{+}$ ground state. (We use here the notation relative to Hund's coupling case {\it c} because spin-orbit coupling in Xe is strong \cite{herzberg}.)

 By contrast, for $P\geq50\,$kPa, collisions rapidly quench the vibrational excitation of excimers and overlapping bound--free emissions from the lowest vibrationally relaxed $(0_{u}^{-}, 1_{u})$ molecular states correlated with the metastable 6s$\,(^{3}\!P_{2})$ atomic state produce the 2nd continuum \cite{dutuit1978,wenck1979}. 

The appearance of the two continua in different pressure ranges is explained in terms of the different kinetics of the processes leading to excimer formation and decay. These processes have been investigated and identified in several spectral- and time-resolved experiments in which lifetimes and rate constants are determined in the low pressure range ($P\leq 0.1\,$MPa) \cite{gornik1982,leichner1976,broadmann1977,millet1978,wenck1979,bonifield1980,raymond1984,salamero1984,moutard1986,moutard1988,museur1994,alekseev1999,ledru2006}. 

The structure of the lowest lying excited molecular states has been theoretically determined either with {\it ab initio}- \cite{ermler1978,jonin2002_II} or model \cite{mulliken1970} calculations, and experimentally by analyzing spectroscopic data \cite{gornik1982,keto1979,raymond1984,castex1981,koe1995}. 

In spite of these efforts, the high pressure ($P> 0.1\,$MPa) features of the VUV continua in the gas have not been extensively investigated  \cite{arai1969,koe1974,lorents1976,moutard1988,morikawa1989} in spite of the clear  connection with the problem of excimer luminescence \cite{chesno1972,saxton1974,schuessler2000,bressi2000,petterson2002} and exciton formation and emission \cite{jortner1965,rice1966,raz1970,suemoto1977} in the condensed phase, which are of both practical and fundamental interest.

It has to be noted that, in addition to this (relative) lack of interest in the high-pressure features of the VUV excimer luminescence, practically no experiments are aimed at investigating the possibility that molecular transitions lying in the infrared (IR) range may occur in the cascade of processes leading to the formation of the VUV radiating states. In particular, as far as emission processes in Xe are concerned, there is only scant indication that a broad IR emission spectrum occurs in the range 765--830 nm \cite{richmann1993}, or, more precisely, in the range 780--815 nm \cite{keto1997}. This spectrum is attributed to a bound--bound transition from vibrational levels of the molecular $0_{g}^{+}$ state, correlated with the 6p$\,[1/2]_{0}$ atomic limit (in Racah notation), toward the $B\,0_{u}^{+}$ (6s$\,[3/2]_{1}$)$_{v^{\prime}\gg0}.$

The absence in literature of further mesurements of IR molecular emission spectra might be ascribed to the fact that the potential energy minimum  of higher lying bound excimer states occurs for an internuclear distance at which the weakly bound ground-state potential is strongly repulsive and, thus, these molecular states are not easily reached by multiphoton selective excitation. 

On the other hand, broad-band excitation accomplished with high-energy charged particles \cite{leichner1976,davide1} may produce excited atoms with such high kinetic energy that can collide at short distance with ground-state atoms, thus yielding higher lying excimer states, though without any control over their parity.

Actually, we have recently observed for the first time a broad IR molecular emission spectrum peaked at about $\tilde \nu_{\mathrm{m}} \approx 7800\,$cm$^{-1}$ $(\lambda_{\mathrm{m}}\approx 1.3 \,\mu\mathrm{m})$ in both pure Xe gas and in a Xe(10\%)--Ar(90\%) mixture in the quite wide pressure range $0.01\,\mathrm{MPa}<P<0.9\,\mathrm{MPa}$. The luminescence is obtained by exciting the gas at room temperature with 50 ns-long pulses 
of 70 keV electrons. Details can be found in literature \cite{Borghesani2001}.

The lR emission spectrum in both of the investigated gases is believed to be produced by the deexcitation of the same molecular species, namely Xe$_{2}.$ Actually, in the Ar--Xe mixture, energy transfer from excited Ar atoms to radiative states of Xe occurs readily and the excited Xe$_{2}$ dimer is produced in heteronuclear three-body reactions of the type
\[ \mathrm{Xe}^{\star} + \mathrm{Xe} + \mathrm{Ar} \to \mathrm{Xe}_{2}^{\star} + \mathrm{Ar}\]
where $^{\star}$ labels an excited state.

The features of this IR continuum are very similar to those of the 2nd VUV continuum. At low pressure, $P\approx 20\,$kPa, its full width at half maximum (FWHM) is $\mathit{\Gamma} \approx 900\,$cm$^{-1}$ and the ratio of $\mathit{\Gamma}$ to $\tilde\nu_{\mathrm{m}}$ is $\mathit{\Gamma}/\tilde\nu_{\mathrm{m}}\approx 0.115,$  to be compared with the value 0.116 of the 2nd VUV continuum \cite{koe1974,jonin1998}.

Owing to these similarities between the IR and VUV continua, we have attributed the IR emission spectrum to a transition between one bound and one dissociative level of the Xe$_{2}$ excimer which are lying higher in the energy relaxation pathway that eventually leads to the population of the excimer levels responsible of the emission of the VUV continua. 

The study of the spectrum as a function of pressure has led to the discovery that the center of the band is red-shifted as the gas density increases. $\tilde \nu_{\mathrm{m}}$ decreases linearly with increasing the gas density $N$ up to a value $N\approx 2\cdot 10^{26}\,$m$^{-3}=7.5\, N_{\mathrm{ig}},$ where $N_{\mathrm{ig}} =2.65\cdot 10^{25}\,$m$^{-3}$ is the density of the ideal gas at $T=273.15\,$K and $P=0.1\,$MPa. The red-shift is larger in the pure gas than in the mixture.  

The red-shift of the emission spectrum has been explained by taking into account the action of the dense environment on the excimer. The electronic structure of a homonuclear RG excimer can be described by an ionic core surrounded by an electron in a diffuse Rydberg-type orbital much larger in diameter than the internuclear distance \cite{mulliken1970,arai1978,audouard1991}. 

It has been suggested \cite{Borghesani2001} that the higher-lying Rydberg-like levels of the excimer are influenced by two density-dependent effects. The first one is the dielectric screening of the Coulomb electron--nucleus interaction 
because of the presence of atoms of the host gas enclosed within the electron orbit. This solvation effect leads to a reduction of the difference between the different electron energy levels, thus always producing a red-shift of the emission spectrum. 

The second effect is a consequence of the wave nature of the electron. The wavefunction of the electron in the Rydberg-like orbital is quite delocalized and can be considere similar to that of a quasifree electron moving through the gas.
Thus, the wavefunction spans a region containing several gas atoms with which it interacts simultaneously. Multiple scattering then leads to a density-dependent shift of the electron energy that is positive or negative depending on the repulsive or attractive character of the electron--atom interaction \cite{fermi1934}.

This physical picture may retain its validity in a high-pressure environment provided that the optically active electron is only weakly scattered off the gas atoms. This situation surely occurs at low pressures, up to a few MPa,  for which the electron mean free path is larger than the orbit of the Rydberg state \cite{huang1978}.

 It is, therefore, interesting to investigate the evolution of a molecular Rydberg state as the density of the environment is progressively increased. This investigation might yield useful pieces of information about the changes of the molecular properties of Wannier--Mott-type impurity states in high-density liquids which have a unique parentage to the Rydberg states of an isolated molecule in a low-density gas \cite{raz1970,messing19771,messing19772} and it also may help understanding the reasons why no IR fluorescence has been 
detected yet in liquefied RG's except superfluid He II \cite{dennis1969,hill1971,keto1974b}.

We have therefore extended the range of  the previous measurements \cite{Borghesani2001,Borghesani2005,borghesani2007ps} to higher pressures, up to $3.3 \,$MPa, corresponding to a gas density $N=8\cdot 10^{26}\,$m$^{-3}= 30\, N_{\mathrm{ig}},$ in order to test the validity of the model developed for lower densities.

\section{Experimental details }\label{sec:expdet}
The experimental setup has been described in a previous paper \cite{Borghesani2001}.
We briefly recall here the main features of the apparatus and the changes we have adopted in order to improve and extend the previous measurements.

A home-made electron gun is used to inject electrons into the sample cell in order to excite the gas. The chopped light beam of a Ne--Ar--F excimer laser (Optex, Lambda Physics) impinges on a gold-coated brass photocathode with a repetition rate as high as 200 Hz. The photoelectrons can be accelerated up to 100 keV by means of a d.c. HV supply (Spellman, mod. RHR120W). Typically, the electron energy is set at 70 keV. 

The 50-ns long electron bunch produced by the electron gun enters the sample cell through a 25-$\mu$m thick Kapton window. The window is thin enough that a significant amount of electrons enters the cell with an acceptable energy degradation (10 to 20 keV of energy are lost in crossing the window), yet it is thick enough to withstand gas pressures as high as 4 MPa. The amount of charge injected into the gas, $0.1<Q<10\,$nC, depends on several factors including the aging of the photocathode of the electron gun, on the optimization of the magnetic deflection and focusing lenses, and so on. 

The sample cell is kept at room temperature. Before being filled with the gas, it is heated up to $\approx 250\,^{\circ}$C and evacuated down to approximately $10^{-5}\,$mbar for several hours. The research grade gas is passed through an Oxisorb cartridge inserted in the filling line to increase its purity. A final impurity content of the gas is estimated to be $\leq 1\,$ppm. The gas pressure is measured by means of a calibrated Sensotec pressure gauge (STJE/1833-35). The gas density is computed by using the appropriate equations of state of Xe and Ar \cite{rabinovich1988}. The law of ideal gas mixtures is used to compute the density of the Ar--Xe mixture \cite{guggenheim}. 

The IR light produced by the Xe$_{2}$ excimer deexcitation exits the cell through a quartz window and is analyzed by a FT--IR (Fourier Transfom InfraRed) Michelson-type interferometer (Equinox 55, Bruker Optics). The interferometer is operated in stepscan  mode \cite{palmer1989} because the source is pulsed. For each mirror position, the signals produced by 50 to 100 excitation events are accumulated and averaged together in order to reduce the effect of the fluctuations of the source intensity.

The interference signal is detected by an InGaAs photodiode (EG\&G, mod. G5832-05) whose output is preamplified by an active integrator of 0.25 mV/fC gain and 400 $\mu$s integration constant. The InGaAs detector shows flat responsivity in the spectral range $(0.6\leq \tilde \nu\leq1.2)\cdot 10^{4}\,$cm$^{-1}$ that is adequate for low to moderate densities $N< 4\cdot 10^{26}\,$m$^{-3}.$ 

For larger densities, the density-dependent red-shift is so large that the excimer band moves outside the range of the detector. Therefore, we have extended the range to be investigated by using a liquid N$_{2}-$cooled InSb photodiode detector (EG\&G, mod. J10D--M204--R02M) whose sensitivity extends farther down to approximately $1800\,$cm$^{-1}.$ In this case, a transimpedance amplifier (EG\&G, mod. PA--9) is used because of the very large output capacitance of the InSb photodiode. 

The output signal is fed to a shaper amplifier (EG\&G, mod. Ortec 575A) with 3 $\mu$s shaping time in order to reduce electrical noise. Finally, the output signal is digitized by a a 16-bit A/D converter and stored in a personal computer for offline analysis. Several spectra, typically ten of them, are recorded for each density and averaged together in order to further improve the quality of the measurement. All of the spectra shown in the next pictures are thus averages over 10 single spectra.

\section{Experimental results and discussion}\label{sec:res}
In this section we present the experimental results. Although the present paper deals with the behavior of the IR excimer continuum as a function of the density, especially at higher densities than previoulsy measured, nonetheless it is worth recalling here briefly the results obtained at low density \cite{borghesani2007jpb}. Actually, the recent publication of accurate potential energy curves for Xe excimer states higher in energy than the first ones \cite{jonin2002_II} has allowed to unequivocally assign the molecular states involved in the deexcitation process investigated in this paper by analysing the spectrum recorded at low density. This piece of information is required in the following when the model for the density red-shift is developed. Thus, we first show and discuss the spectrum recorded at low density.

\subsection{IR emission spectrum at low density: Identification of the molecular states}\label{subsec:lowN}

\begin{figure}[htbp]
 \begin{center}
 \includegraphics[angle=90,width=\columnwidth]{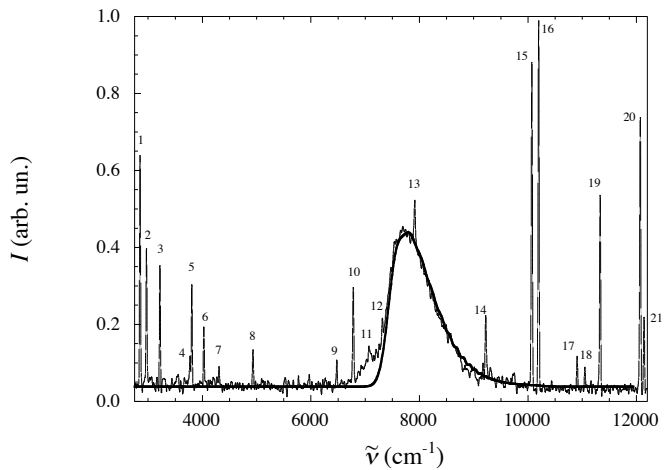}
\caption{\small High-resolution IR emission spectrum of Xe$_{2}$ excimers in pure Xe gas for $T\approx 300\,$K and $P\approx 0.1\,$MPa. Xe I lines are numbered. Thick line: Franck-Condon calculation.\label{fig:spclowN}}
 \end{center} \end{figure}

The IR emission spectrum recorded in pure Xe gas at room temperature for a pressure $P\approx 0.1\,$ MPa is shown in Fig. \ref{fig:spclowN}. The corresponding gas density is $N\approx 2.45 \cdot 10^{25}\,$m$^{-3}.$ 
 A detailed analysis of this spectrum can be found in literature \cite{borghesani2007jpb}. 
Here, we recall the main features of the spectrum and sketch a brief outline of its analysis by quoting the main results.
 
 This spectrum has been recorded at high resolution (16 cm$^{-1}$) in order to detect atomic lines that help ascertain the relationship between the excimer states and their parent atomic states.
 
The excimer contribution to the spectrum is the continuous structure displayed in the center of Fig.\ref{fig:spclowN} and is surrounded by several Xe I atomic lines that can be easily identified by inquiring the database of atomic spectra at the NIST website \cite{NIST}.
The assignment of the atomic Xe I lines is summarized in Tab. \ref{tab:correspondence} in which the electronic states involved in each transition are reported along with the proper reference.
\begin{table}
\caption{\label{tab:correspondence}\small Xe I lines assignment}
\begin{ruledtabular}
\begin{tabular}{lcr}
Line&Electronic states&Reference\\
\hline
  1 & 6p--5d &\cite{humphreys1973}  \\
  2 & 6p--5d  &\cite{humphreys1973}  \\
  3 & 6p--5d  &\cite{humphreys1973}  \\
  4 & 6p--5d  &\cite{hepner1961}  \\
  5 & 6p--5d  &\cite{hepner1961}  \\
  6 & 6p--5d  &\cite{hepner1961}  \\
  7 & 6p--5d  &\cite{hepner1961}  \\
  8 & 6p--5d  &\cite{mishra2000}  \\
  9 & 7s--6p  &\cite{humphreys1952}  \\
  10 & 7s--6p  &\cite{humphreys1952}  \\
  11 & 7s--6p&\cite{humphreys1952}  \\
  12 & 7s--6p  &\cite{humphreys1973}  \\
  13 & 7s--6p  &\cite{meggers1935}  \\
 14 & 6p--6s  &\cite{meggers1935}  \\
 15 & 6p--6s  &\cite{meggers1934}  \\
 16 & 6p--6s  &\cite{meggers1934}  \\
  17 & 6d--6p, 6p--6s  &\cite{humphreys1933}  \\
 18 & 6p--6s  &\cite{meggers1934}  \\
 19 & 6p--6s  &\cite{meggers1934}  \\
 20 & 6p--6s  &\cite{meggers1934}  \\
 21 & 6p--6s  &\cite{meggers1934}  \\
\end{tabular}
\end{ruledtabular}
\end{table}

Lines 1 through 8 are emitted in transitions between excited 6p and 5d electronic states of atomic Xe. In particular, line 1 is the very well known Xe I line appearing at $\lambda =3.51\,\mu$m \cite{vallee1981}. Lines 9 through 13 stem from 7s--6p transitions. These are the lines closest to the excimer band. Line 13 quite overlaps the excimer spectrum.
Lines 14--16 are due to 6p--6s transitions, whereas line 17 is an unresolved 6p--6s and 6d--6p doublet. Finally, lines 18 through 21 are  related to 6p--6s transitions.
 
 It appears evident from the low density spectrum that atomic states in the 6p manifold are efficiently populated in the energy degradation processes occurring in the gas after excitation by electron impact. The crowding of lines related to atomic 7s--6p transitions in the region of the excimer band suggests that the upper bound molecular state is related to the 6p atomic limit. In particular, line 13 stems from a transition from the atomic 7s$\,[1/2]^{\mathrm{o}}$state to the 6p$\,[1/2]_{0}$ state, thus leading us to reasonably assume that the bound excimer state is related to this atomic limit. 
 
 The experimental observation of three-body quenching of the 6p$\,[1/2]_{0}$ state confirms the hypothesis that a molecular bound state can be formed from this atomic limit by three-body collision-induced association \cite{bowering1986}. The molecule can then predissociate
toward the 5d$\,[1/2]_{1}$ atomic state with a characteristic time $\tau_{p}\approx 10^{-10}\,$s \cite{dehmer1986,museur1994}, but the mean time between collisions at the density of the experiment is so short $\tau_{c}\leq 10^{-11}\,$s \cite{Borghesani2001} as to lead to a quick electronic relaxation and stabilization of the excimer.  

Moreover, the excimer population rapidly attains thermal equilibrium with the surrounding gas because the excimer lifetimes for radiative decay $\tau_{1}$ and for vibrational relaxation $\tau_{2}$ are estimated to be much longer than the mean time between collisions. Actually, though unknown for the state observed in the present experiment, such lifetimes can be assumed to have values similar to those of the VUV radiating states, namely, $\tau_{1}\approx 5\,$ns for the $0_{g}^{+}$ \cite{moutard1988} state and $\tau_{1}\approx 40\, $ns  for the $(0_{u}^{-}, 1_{u})$ states \cite{keto1979}, whereas the decay rate for vibrational relaxation $k_{2}=6.47 \cdot 10^{-17}\, $m$^{3}/$s \cite{salamero1984} yields a lifetime $\tau_{2}\approx 0.65\,$ns at the density $N=2.45\cdot 10^{25}$ at which the spectrum reported in Fig. \ref{fig:spclowN} is measured. 

The identification of the molecular states involved in the bound--free transition yielding the observed spectrum can be accomplished by carrying out traditional Franck--Condon calculations \cite{herzberg}.

Recently, Jonin {\it et al}.  published theoretical potential energy curves for higher excited excimer states \cite{jonin2002_II}.
The choice of the ungerade $(3)0_{u}^{+}$ state, correlated with the 6p$\,(^{1}\!D_{2}) $ atomic  limit, as the bound molecular state, and of the gerade $(1)0_{g}^{+}$ state, correlated with the 6s$\,(^{3}\!P_{1})$ atomic limit,  as the dissociative state, fulfills all the required selection rules and allows to simulate a spectrum shape that neatly fits the experimental one \cite{borghesani2007jpb}. Henceforth, let us term $V_{u}$ and $V_{g}$ the potential energy of the upper and lower states, respectively.

In the centroid approximation \cite{herzberg,telli}, the spectrum intensity is given by 
\begin{eqnarray}
 I& \propto &\sum \limits_{v^{\prime} J^{\prime}}\mathrm{e}^{-\beta E_{v^{\prime}\!J^{\prime}}} \left\{ \left( J^{\prime} + 1 \right) \Big\vert\langle\epsilon^{\prime\prime},J^{\prime} + 1 \vert v^{\prime},J^{\prime}\rangle\Big\vert^{2} \right.
 \nonumber \\ 
 & +&\left. 
 J^{\prime}
  \Big\vert\langle\epsilon^{\prime\prime},J^{\prime}-1\vert v^{\prime},J^{\prime}\rangle\Big\vert^{2}\right\}\tilde\nu^{4}
  \label{eq:fcspc}\end{eqnarray}
Primed quantities refer to the upper bound state whereas doubly primed ones refer to the lower dissociative state. $\beta^{-1} = k_{\mathrm{B}}T,$ where $k_{\mathrm{B}}$ is the Boltzmann constant and $T$ is the gas temperature.
$E_{v^{\prime}\! J^{\prime}}$ are the rovibrational energy eigenvalues of $V_{u}.$   $\vert v^{\prime},J^{\prime}\rangle $ are the rovibrational eigenfunctions of $V_{u},$ described by the vibrational and rotational quantum numbers $v^{\prime}$ and $J^{\prime},$ respectively. $\vert\epsilon^{\prime\prime},J^{\prime\prime}\rangle$ are scattering states of kinetic energy $\epsilon^{\prime\prime}$ and angular momentum $J^{\prime\prime}$ in the vibrational continuum of the dissociative potential $V_{g}.$
$\tilde \nu$ is the emission wavenumber, i.e., $hc\tilde \nu$ is the energy of the photons emitted in the electronic transition from $V_{u}$ to $V_{g}.$ As usual, $h$ and $c$ are the Planck's constant and the speed of light, respectively.
In Eq. \ref{eq:fcspc} the selection rule $J^{\prime} -J^{\prime\prime}=\pm 1$ has been used. Details of the numerical calculations can be found in literature \cite{borghesani2007jpb}.

In Fig. \ref{fig:spclowN} the simulated spectrum shape is represented by a thick solid line. The good agreement between experiment and numerical simulation lends credibility to the proposed assignment of the molecular states. 

Although several rovibrational of $V_{u}$ are thermally excited, nonetheless the main contribution to the spectrum intensity comes from the rovibrational ground state. Thus, roughly speaking, the central wavenumber $\tilde\nu_{\mathrm{m}}$ of the spectrum corresponds to the difference $V_{u}-V_{g}$ averaged over the spatial extension of the wavefunction of the vibrational ground state of $V_{u}$
\begin{equation}
\tilde \nu_{\mathrm{m}} \simeq \langle \left(V_{u}-V_{g}\right)\rangle/hc
\label{eq:numvave}\end{equation}
This piece of information will be used in the following to intepret the density dependence of the spectrum features.
 \subsection{Density dependent red--shift of the spectrum}\label{sec:alienati}
\begin{figure}[htbp]
 \begin{center}
 \includegraphics[angle=270,width=\columnwidth]{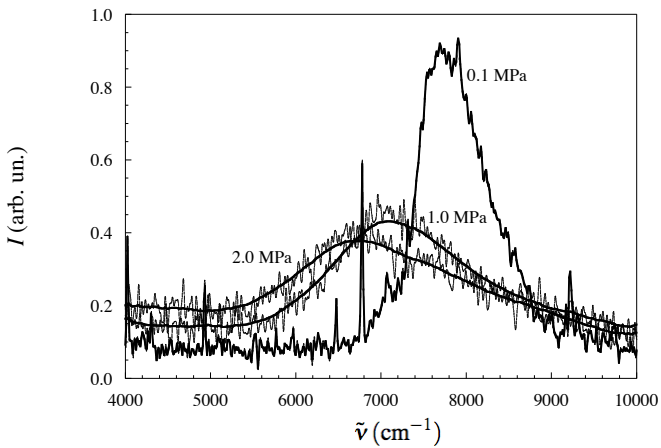}
\caption{\small High-resolution ($16\,$cm$^{-1}$) spectra recorded in pure Xe gas for $P=0.1,$ 1, and $2\,$MPa. The thick solid lines are interpolating curves drawn for visual help. \label{fig:XEN}}
 \end{center} \end{figure}

As the gas pressure is increased, the IR emission spectrum of the excimer Xe$_{2}$ changes drastically. 
In Fig. \ref{fig:XEN} three spectra recorded in pure Xe gas for $P=$0.1, 1, and $2\,$MPa, respectively, are shown. 
 
\noindent Except for the disappearance of some atomic lines, two main effects can be easily observed  as a consequence of the pressure increase. The first one is a strong red-shift of the center of gravity of the spectrum, whereas the second one is its broadening. In pure Xe gas, shift and broadening are so large as to make the excimer spectrum increasingly overlap the group of atomic lines numbered as 1 through 8 of Fig. \ref{fig:spclowN}. Moreover, for $P\geq 0.8\,$MPa, approximately, the InSb detector is mandatorily required in order to record the spectrum in its entirety.

\begin{figure}[!t]
 \begin{center}
 \includegraphics[angle=270,width=\columnwidth]{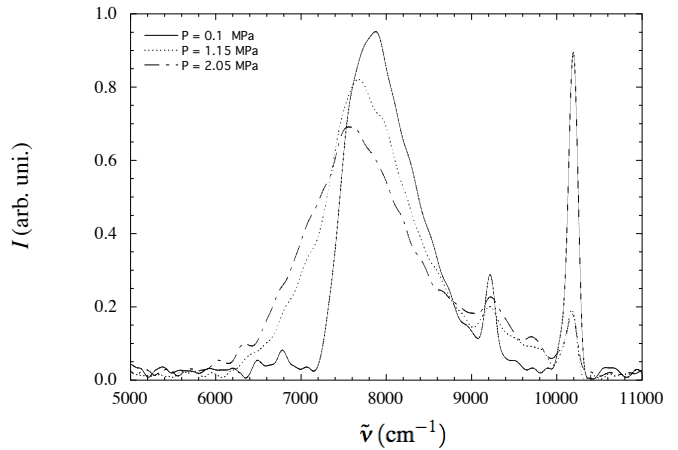}
\caption{\small Spectra recorded in the Ar(90\%)--Xe(10\%) mixture for $P=0.1,$ 1.15, and $2.05\,$MPa. The resolution is 100 cm$^{-1}.$ \label{fig:ARXEN}}
 \end{center} \end{figure}

The same qualitative effects occur in the Ar--Xe mixture although their entity is much smaller, as can be observed in Fig. \ref{fig:ARXEN}. Also in this case, the spectrum shifts to the red side and broadens with increasing gas pressure. However, the amount of shift and broadening is far smaller than in the case of pure Xe.
The three spectra reported in this figure can be directly compared with those shown in Fig. \ref{fig:XEN} because the corresponding pressures are nearly the same. The only difference between the spectra of Fig. \ref{fig:XEN} and Fig. \ref{fig:ARXEN} is the experimental resolution, $16\,$ cm$^{-1}$ in the former figure, $100\,$cm$^{-1}$ in the latter one.

Thus, the effect of a changing environment is very evident. Whereas the increase of density is expected to result in a higher collision frequency that reduces the lifetime of the excimer species and increases the width of the continuum, such large a density dependent red-shift of the spectrum is not expected on the basis of the theories of pressure broadening of atomic lines in the impact approximation \cite{margenau1936,jablonski1945,foley1946,anderson1949,chen1957,baranger1958c,byron1964}. At the same time, the red-shift difference between the pure gas- and the mixture case is hardly explained at all.

\begin{figure}[htbp]
 \begin{center}
 \includegraphics[angle=270,width=\columnwidth]{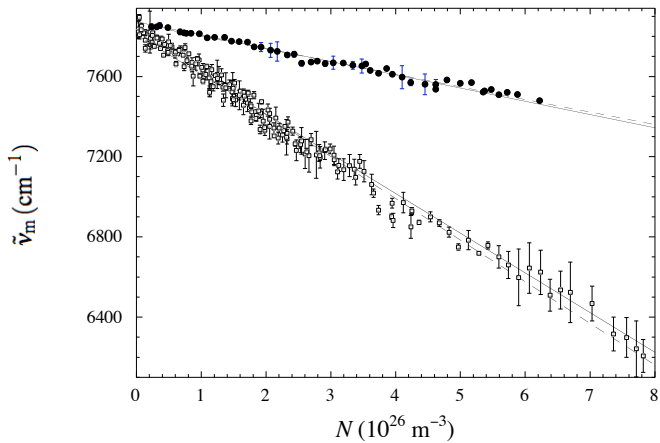}
\caption{\small $\tilde\nu_{n} $ vs $N.$ Closed points: Ar--Xe mixture. Open points: pure Xe gas. Dashed lines: linear fit to the data. Solid lines: prediction of the theoretical model developed in the text.\label{fig:XeArXeshift}}
 \end{center} \end{figure}

In this section we will deal with the density dependent red-shift. The wave number of the maximum of the excimer band, $\tilde \nu_{\mathrm{m}},$ 
is used for its quantitative characterization. $\tilde \nu_{\mathrm{m}}$ has been determined by locally fitting a cubic polynomial to the spectrum near its maximum. A cubic polynomial has been chosen instead of a parabola in order to better locate the spectrum maximum because of the slight asymmetry of the spectrum shape in its neighborhood. $\tilde\nu_{\mathrm{m}}$ is plotted as a function of the gas density $N$ in Fig. \ref{fig:XeArXeshift}.

The error bars on the data points are an estimate of the uncertainty with which the values $\tilde\nu_{\mathrm{m}}$ are determined by the fitting procedure we adopted.

 $\tilde \nu_{\mathrm{m}}$ strongly decreases to smaller wavenumber values with increasing $N$ in both the pure gas and in the mixture though the effect, as anticipated, is far stronger in the pure gas than in the mixture. The density dependence of $\tilde\nu_{\mathrm{m}}$ is linear in both types of gas up to the highest investigated density $N=8\cdot 10^{26}\,$m$^{-3}=30 N_{\mathrm{i.g.}}.$ The present measurements more than double the density range explored in the previous experiments \cite{Borghesani2001,Borghesani2005}. In the region for $N\leq 4 \cdot10^{26},$ common to all of the experiments, the present data agree perfectly with the previous ones. The latter have not been shown in the picture only for the sake of graphical clarity.

The behavior of $\tilde \nu_{\mathrm{m}}$ as a function of the density $N$ can accurately be described by the linear relationship
\begin{equation}
\tilde\nu_{\mathrm{m}}= \tilde\nu_{\mathrm{m},0} -A N
\label{eq:fitlin}\end{equation}
Such linear fit is represented by dashed lines in Fig. \ref{fig:XeArXeshift} and
the values of the fitting parameters are reported in Tab. \ref{tab:fitpar}
\begin{table}[!b]
\caption{\label{tab:fitpar}\small Fitting parameters of the red-shift}
\begin{ruledtabular}
\begin{tabular}{lcr}
Gas&$\tilde\nu_{\mathrm{m},0}$ (cm$^{-1}$)&A ($10^{-22}\,$m$^{2}$)\\
\hline
  Xe & $7829 \pm 26$ & $2.08\pm 0.04 $ \\
 Ar--Xe & $7864\pm 13$ &$ 0.63\pm0.02$\\
\end{tabular}
\end{ruledtabular}
\end{table}

 Within the experimental accuracy the zero-density limit of $\tilde\nu_{\mathrm{m}}$ is the same in both the pure gas and the mixture. This means that the molecular species Xe$_{2},$ which is responsible for the IR emission in the pure gas, is also produced in the mixture. This fact is not unexpected as it is well known that excitation is readily transferred from the species of larger ionization potential (Ar) to the species of smaller ionization potential (Xe) \cite{galy1992}. In this way, the electrons injected into the mixture primarily excite Ar atoms that transfer their excitation to Xe atoms. Then, though low, the concentration of Xe atoms is sufficiently high to make their 
collisions with ground-state Xe atoms highly probable, thus leading to a relevant production of the Xe$_{2}$ excimer.

The most striking feature exhibited by $\tilde\nu_{\mathrm{m}}$ is its large negative slope as a function of the density whose absolute value is a couple of orders of magnitude larger than the density dependence of the shift (either blue or red) of atomic lines \cite{laporte1975,raymond1984,moutard1986,alford1992}. The slope in the pure gas is more than three times larger than in the mixture. 

These experimental observations can be rationalized by a simple model previously developed by us \cite{Borghesani2001} in which two differents effects are taken into account. The description of both of them is based on the assumption that the electronic structure of an excited state of a homonuclear diatomic molecule is quite accurately described by an ionic core and an electron in a diffuse Rydberg orbital much larger than the internuclear distance \cite{mulliken1970}. Such a state can exist also in a high-pressure environment provided that the Rydberg electron is only weakly scattered off the atoms of the host gas. This condition is fulfilled for the pressures of the present experiments because the mean free path of quasifree electrons is several nanometers long \cite{huang1978}, i.e., much longer than the radius of the orbit of the Rydberg electron. 
\paragraph{Dielectric screening, or solvation, effect}
The first effect is the dielectric screening of the Coulomb interaction between the electron and the ionic core because the molecule is immersed in a dielectric medium. It is assumed that the Rydberg electron is so greatly delocalized that many atoms of the host gas are encompassed within its orbit. They are polarized by the strong field of the ionic core and, hence, they reduce the Coulomb interaction between electron and core. If the gas density is large enough the dielectric screening effect can be treated within a continuum approximation through the use of the (density dependent) dielectric constant $K(N)$ of the medium. 

It has been shown in the case of the hydrogen molecular ion H$_{2}^{+}$ \cite{herman1956} that the electron energy eigenvalues, which can be considered as the potential energy curves for the nuclear motion in the Born--Oppenheimer approximation \cite{haken,atkins}, are reduced by a factor $K^{-2}$ if the ion is immersed in a dielectric medium. This result can be easily generalized for any interaction potentials of electrostatic nature acting on the optically active electron that can be expressed in terms of a multipolar expansion.

 In particular, in the present case, the potential energy curves of the initial (bound) molecular state $(3)0_{u}^{+},$ $V_{u},$ and of the final (dissociative) state $(1)0_{g}^{+}, $ $V_{g},$ must be divided by the square of the dielectric constant. Hence, also their difference is reduced by the same factor. 
 
 As a consequence of the dielectric screening, the wavenumber of the photon emitted in the transition between the initial and final electronic states of the excimer is lowered as the density is increased
 \begin{equation}
\tilde\nu_{\mathrm{m}}= \frac{\langle \left(V_{u}-V_{g}\right)\rangle}{hcK^{2}\left(N\right)}\approx
\tilde\nu_{\mathrm{m},0}\left(1-\frac{2\alpha}{\epsilon_{0}}N\right)
\label{eq:solveff}\end{equation}
where $\alpha$ is the atomic polarizability of the atoms of the host gas. $\alpha = 4.45\cdot 10^{-40}\,$F$\cdot$m for Xe and  $\alpha = 1.827\cdot 10^{-40}\,$F$\cdot$m for Ar \cite{maitland}. $\epsilon_{0}$ is the vacuum permittivity. $K$ can be calculated as a function of $N$ by using the Lorentz--Lorenz formula $(K-1)/(K+2)=(\alpha/3\epsilon_{0})N.$ Finally, in Eq. \ref{eq:solveff} we have Taylor expanded $K$ to leading order in $N$ with a relative error $<0.04\% $ at the highest density.

$\tilde\nu_{\mathrm{m},0}\approx 7800\,$ cm$^{-1}$ is determined by a linear fit to the data.
Thus, the contribution of this solvation effect to the slope of $\tilde\nu_{\mathrm{m}}$ amounts to $2\tilde\nu_{\mathrm{m},0}\alpha/\epsilon_{0}\approx 7.8\cdot 10^{-23}\,$m$^{2}$ for Xe and to $3.2\cdot 10^{-23}\,$m$^{2}$ for the mixture.  Note that, in the case of the Ar--Xe mixture, we have used the value of polarizability 
relative to Ar. Actually, the Xe concentration in the mixture $(\approx 10\,\% )$ is high enough for allowing the formation of the excimer but is sufficiently low for the newly formed Xe$_{2}$ species to be surrounded, on average, by Ar atoms only.

For both gases, this contributiondue to dielectric screening only accounts for approximately 50\% or less of the observed slope. The rest can be accounted for by including the quantum multiple scattering effect.

\paragraph{Quantum multiple scattering effect}
The second, density-dependent effect that influences the slope of the center of gravity of the spectrum is a quantum multiple scattering effect. Its quantitative description is due to Fermi \cite{fermi1934} who explained the density-dependent shift of the high spectroscopic terms of the absorption lines of alkali vapors in a buffer gas \cite{amaldi1934}. 

The optically active electron is assumed to be loosely bound to the ionic core and it is treated as if it were moving through the gas like a quasifree electron. Its wavefunction is thus quite delocalized and spans a region containing several atoms of the surrounding gas. 

In this situation, the simultaneous interaction of the electron with many atoms gives origin to multiple scattering effects that lead to a density-dependent shift $V_{0}(N)$ of the electron ground-state energy. The shift depends on the electron--atom scattering properties, namely, the scattering length $a,$ and for not too high densities is proportional to the density itself
\begin{equation}
V_{0}=\frac{2\pi\hbar^{2}}{m}Na
\label{eq:v0}\end{equation}
Here, $m$ is the free electron mass and $\hbar$ is the reduced Planck's constant. More recent calculations have confirmed the validity of the Fermi's result up to moderate values of $N$ \cite{springett1968,plenckiewicz1986,boltjes1993}.

This contribution must clearly be added only to the electronic energy of the upper bound molecular state because the electronic wavepacket necessarily collapses on either  atoms of the molecule during the transition followed by dissociation, Hence, as a function of $N,$ the electronic energy of the bound state changes in the following way
\begin{equation}V_{u}\to \frac{V_{u}}{K^{2}(N)}+V_{0}(N)
\label{eq:vuff}\end{equation}
If the electron--atom interaction is attractive, $a<0$ and $V_{0}<0,$ leading to a red-shift of the spectrum. The opposite behavior (a blue-shift) is shown by gases whose interaction with the electron is repulsive, yielding $a>0$  and $V_{0}>0$ \cite{reininger1982,reininger1983,kohler1987}. 

We are finally led to the following expression for the energy of the IR photon emitted in the center of the spectrum in the transition between the two electronic states of the molecule
\begin{eqnarray}
 \tilde\nu_{\mathrm{m}}& =&\frac{\langle\left(V_{u}-V_{g}\right)\rangle}{hcK^{2}(N)} + \frac{V_{0}(N)}{hc} \nonumber\\ 
  &\approx  &\tilde\nu_{\mathrm{m},0}\left( 1 - \frac{2\alpha}{\epsilon_{0}}N\right) + \frac{\hbar}{mc}Na\nonumber\\
 &=&\tilde\nu_{\mathrm{m},0} - \left(  \tilde\nu_{\mathrm{m},0} \frac{2\alpha}{\epsilon_{0}} -  \frac{\hbar a}{mc}
 \right)N
 \label{eq:cicci}\end{eqnarray}
This model predicts a (negative) linear density dependence of $\tilde\nu_{\mathrm{m}},$ as experimentally observed up to $N=8\cdot 10^{26}\,$m$^{-3},$ with theoretical slope $A_{\mathrm{th}}$
given by
\begin{equation}
A_{\mathrm{th}}=\left(  \tilde\nu_{\mathrm{m},0} \frac{2\alpha}{\epsilon_{0}} -  \frac{\hbar a}{mc}
 \right)
\label{eq:}\end{equation}
By using literature values of the scattering length, namely $a=-3.09 \cdot 10^{-10}\,$m for Xe and, for the reasons previously explained, $a=-0.86\cdot 10^{-10}\,$m for Ar \cite{zecca}, we obtain the values $A_{\mathrm{th}}= 1.98\cdot 10^{-22}\,$m$^{2}$ for Xe and $A_{\mathrm{th}}= 0.66\cdot 10^{-22}\,$m$^{2}$ for the Ar--Xe mixture. These values are in very close agreement with those of the linear fit of the experimental data  reported in Tab. \ref{tab:fitpar}. 
The solid straight lines in Fig. \ref{fig:XeArXeshift} represent Eq. \ref{eq:cicci}. This figure confirms the excellent description that the present model gives of the experiment.

It is worth noting that the two contributions to the overall spectral shift may not act in the same direction. Actually, the dielectric screening always reduces the strength of the Coulomb interaction of the Rydberg electron with the ionic core, thus always producing a red-shift of the spectrum. 

By contrast, the shift of the ground-state energy of a quasifree electron, $V_{0},$ may be either negative or positive, depending on the sign of the scattering length. Whereas $V_{0}<0$ in gases for which the electron--atom interaction is attractive, $V_{0}>0$ for repulsive gases, such as Ne and He. 

We, indeed, observe that the Xe$_{2}$ emission spectrum is less red-shifted in Ar than in Xe, mainly because $V_{0}$ in Ar \cite{tauchert1977,borghesani1991} is less negative than in Xe \cite{tauchert1977}.  We thus expect that the shift would practically be absent in a Xe--Ne mixture because the electron--Ne scattering length is small but positive \cite{omalley1980hb}, yielding a positive though small contribution to the electron energy that nearly compensates the red-shift due to dielectric screening. Furthermore, in a He--Xe mixture the electron--He scattering length is so large and positive \cite{omalley1963}, yielding $V_{0}\approx 1\,$eV \cite{sommer1964}, and the atomic polarizability is so small \cite{maitland} that the multiple scattering contribution should overcompensate the dielectric screening, thus producing an overall blue-shift of the excimer emission spectrum. 

In some sense, the observation of the density-dependent shift of the IR excimer emission spectrum could be considered as a technique for measuring the density-dependent shift of the ground-state energy of a quasifree electron in a dense gas in alternative to other methods \cite{reininger1990,reininger1991}.

 \subsection{Density--broadening of the excimer band}\label{sec:brodo}
This section is devoted to the discussion of the other relevant feature exhibited by the excimer emission spectrum as the gas density is increased, namely, the large broadening of the molecular continuum that can be observed in Fig. \ref{fig:XEN} and in Fig. \ref{fig:ARXEN}. 

\begin{figure}[htbp]
 \begin{center}
 \includegraphics[angle=270,width=\columnwidth]{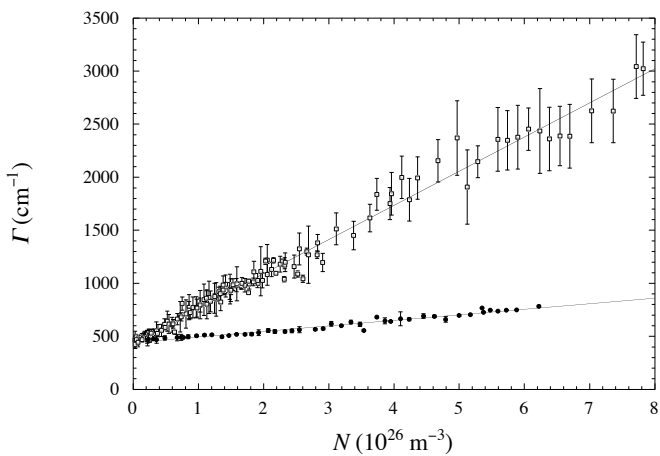}
\caption{\small HWHM $\mathit{\Gamma}$ of the IR excimer emission spectrum as a function of $N.$ Open points: pure Xe gas. Closed points: Ar--Xe mixture.\label{fig:brodo}}
 \end{center} \end{figure}

The spectrum shape is not as simple as that of an atomic line but is rather given by the complicated convolution of the contributions of several Franck--Condon factors expressed by Eq. \ref{eq:fcspc}. Moreover, it is quite asymmetric, though its asymmetry is reduced as the density is increased. For these reasons there is no easy way to quantify the spectrum broadening. 
In spite of this, we characterize it by measuring its half width at half maximum (HWHM), $\mathit{\mathit{\Gamma}},$ as a function of $N.$ 
The raw values of the HWHM directly determined from the spectrum, $\mathit{\mathit{\Gamma}}_{\mathrm{exp}},$ are corrected for the width of the instrumental function, $\mathit{\mathit{\Gamma}}_{\mathrm{ins}},$ in the usual way
\begin{equation}
\mathit{\mathit{\Gamma}} = \left[\mathit{\mathit{\Gamma}}^{2}_{\mathrm{exp}}-\mathit{\mathit{\Gamma}}^{2}_{\mathrm{ins}}\right]^{1/2}
\label{eq:rawgamma}\end{equation}
In Fig. \ref{fig:brodo} we report $\mathit{\mathit{\Gamma}}$ as a function of $N.$

\noindent In the whole density range explored $\mathit{\mathit{\Gamma}}$ is a linear function of $N$ for both types of gas and is well fitted to a straight line
\begin{equation}
\mathit{\Gamma} = \mathit{\Gamma}_{0} + \gamma N
\label{eq:fitbrodo}\end{equation}
The values of the fitting parameters are reported in Tab. \ref{tab:fitbrodopar}.
\begin{table}[!hb]
\caption{\label{tab:fitbrodopar}\small Fitting parameters of the spectrum broadening}
\begin{ruledtabular}
\begin{tabular}{lcr}
Gas&$\mathit{\mathit{\Gamma}}_{0}$ (cm$^{-1}$)& $\gamma$ ($10^{-22}\,$m$^{2}$)\\
\hline
  Xe & $446 \pm 9$ & $3.22\pm 0.03 $ \\
 Ar--Xe & $443\pm 6$ &$ 0.52\pm0.02$\\
\end{tabular}
\end{ruledtabular}
\end{table}

Within the experimental accuracy, the low-density limit $\mathit{\mathit{\Gamma}}_{0}$ is the same in both the pure gas and in the mixture. This observation further confirms the conclusion that the same molecular species is responsible of the IR emission in the two gases. By contrast, the slope $\gamma$ is approximately 6 times larger in the pure gas than in the mixture. 

Three questions, thus, are to be addressed: the absolute values of the width of the IR continuum, its linear dependence on the gas density, and, finally, the difference between the two gases.

What 
can be deduced from Eq. \ref{eq:fitbrodo}? First of all, as the IR continuum is originated in a bound-free transition, $\mathit{\Gamma}_{0}$ is related to the energy range available to the final state. It is roughly determined by the steepness of the potential energy curve of the dissociative state and by the spatial extension of the wavefunction of the vibrational ground state of the bound potential. In this sense, $\mathit{\Gamma}_{0}$ does not give any further pieces of information in addition to those gathered in simulating the spectrum shape at low density. 

More interesting, by contrast, is the linear dependence of the spectrum width on the gas density that suggests that binary collisions are responsible of the increase of the decay rate of the excimer. Pressure broadening theories have been developed in the impact approximation to account for the effect of foreign perturbers on spectral lines \cite{margenau1936,jablonski1945,foley1946,anderson1949,chen1957,baranger1958a,baranger1958b,baranger1958c,byron1964}. All of them lead to the prediction of an increase of the spectral width, $w,$ that is linear in the gas density
\begin{equation}
w=A N{\bar v} \sigma_{0}
\label{eq:w}\end{equation}
where $A$ is a numerical constant of order unity \cite{baranger1958a}. $\bar v \propto T^{1/2}$ is a thermal velocity of the perturbers and $\sigma_{0} $ is the total cross section (or, optical collision cross section). If $N$ is measured in m$^{-3},$ $\bar v$ in m/s, and $\sigma_{0}$ in m$^{2},$ $w$ is expressed in s$^{-1}.$ $N\sigma \bar v$ is the scattering rate.

Even neglecting the fact that these theories are concerned with spectral lines rather than with a continuum, it is not clear if they are applicable in the present case. Actually, they have been derived under the so called impact approximation, according to which the interatomic separation must be much larger than the atomic diameter. Differently stated, this condition means that the mean free time between collisions must be much longer than the time of the collision itself or, equivalently, that the men free path is much longer than the atomic diameter.

In the conditions of the present experiment, the lowest gas density is of the order of $N_{\mathrm{min}}\approx 2\cdot 10^{25}\,$m$^{-3},$ whereas the highest is $N_{\mathrm{max}}\approx 8\cdot 10^{26}\,$m$^{-3}$. 
 Let us assume a typical density of order $\tilde N = 10^{26}\,$m$^{-3}.$ A rough estimate of the average interatomic distance is $d\sim \tilde N^{-1/3}\approx 20\,$\AA, which reduces to $d\approx 10\,$\AA\ at the highest density. $d$ must be compared with the atomic radii of Xe and Ar, and with the size of the Xe$_{2}$ molecule. 
 
 Several values can be found in literature for the hard-sphere radii $\rho_{\mathrm{Xe}},$ $\rho_{\mathrm{Ar}},$ and $\rho_{\mathrm{AX}}$ of the Xe--Xe,  Ar--Ar, and Ar--Xe interaction, respectively. Namely, $(4.26< \rho_{\mathrm{Xe}}<4.42)\,$\AA\  for Xenon \cite{bobetic1983,maitland1973,maitland1973b}, $(3.7<\rho_{\mathrm{Ar}}<3.8) \,$\AA\  for Ar \cite{maitland,maitland1973}, and $(3.98<\rho_{\mathrm{AX}}<4.11) \,$\AA\  for the Ar--Xe pair  \cite{marks1988}. The internuclear distance in the $(3)0_{u}^{+}$ bound potential of the Xe$_{2}$ molecule is $R_{e^{\prime}}\approx 3.23\,$\AA\ \cite{jonin2002_II}. Thus, it is questionable whether such theories might be applicable at the densities of the present experiment because the interatomic distance is comparable with the atomic or molecular radii. 
 
 However, in a less sophisticated approach, we can resort to using the old Lorentz theory \cite{lorentz1906} in which atoms are assumed to behave as hard-spheres. When a perturber collides with a radiating atom, the radiation process is interrupted completely (or the phase is changed by an arbitrary amount \cite{foley1946}), the vibration energy being converted into kinetic energy. This theory actually predicts the same dependence of the broadening on the density as given by Eq. \ref{eq:w}. Thus, we will use Eq. \ref{eq:w} with $A=1.$ Though $\sigma_{0}$ may not necessarily be the kinetic cross section, it is expected to be of the same order of magnitude \cite{margenau1936}.

The broadening of the spectrum expressed in wavenumbers is simply related to the scattering rate in such a way that the slope $\gamma $ is directly related to the scattering cross section as
\begin{equation}
\gamma =  \frac{1}{2\pi c}\sigma_{0} \bar v
\label{eq:gamma}\end{equation}
The thermal velocity is $\bar v = (3k_{\mathrm{B}}T/\mu)^{1/2},$ where $k_{\mathrm{B}}$ is the Boltzmann constant and $\mu$ is the reduced mass of the Xe$_{2}-$Xe or Xe$_{2}-$Ar systems, respectively. The cross section is calculated from the experimental values of the slope $\gamma.$ We note that, owing to the very large mass of Xe$_{2},$ the broadening due to the Doppler effect is estimated to be $\mathit{\Gamma}_{D}\approx 2\cdot 10^{-2}\,$ cm$^{-1}$ in the band center, and is thus completely negligible in comparison to the broadening due to collisions \cite{Borghesani2005}. 
 
In Tab. \ref{tab:crs} we report the values of the cross sections determined for the pure gas and for the mixture for $T=300\,$K along with the values of the hard--sphere radius $\rho = (\sigma_{0}/\pi)^{1/2}. $%

\begin{table}[!b]
\caption{\label{tab:crs}\small Physical parameters of the spectrum broadening}
\begin{ruledtabular}
\begin{tabular}{lcr}
Gas& Xe&Ar--Xe\\
\hline
$\bar v \,$(m/s) & $292$ &$ 465$\\
$\sigma_{0}\,$(nm$^{2}$) & $2080
 $& $211
 $\\
$\rho\,$ (\AA)& 128&41\\
\end{tabular}
\end{ruledtabular}
\end{table}

  It is known that the optical collision cross sections and diameters depend on the nature and strength of the molecular forces, as well as upon the intrinsic properties of the radiating molecule and are usually greater that the kinetic collision cross sections and diameters \cite{crane1963}. Nonetheless, the values of the cross sections and of the hard-sphere radii determined from the observed slopes and reported in Tab. \ref{tab:crs} turn out so large as not to be reasonable for a RG molecule--RG atom collision. 

We further observe that the cross sections deduced from the experiment in the pure gas and in the mixture are very different from each other in spite of the fact that in both case a RG molecule--RG atom collision is occurring and that the values of the hard-sphere radius of Xe and Ar are not this different. The ratio of the cross sections is $S=\sigma_{0}(\mathrm{Xe})/\sigma_{0}(\mbox{Ar--Xe}) =9.84.$ We believe that this difference is a direct consequence of the indistinguishability of identical particles in quantum mechanics, as we will show later.

We first suggest a possible explanation for the large values of the cross sections. We note that the spectrum is given, roughly speaking, by a convolution of the contributions of many vibrational and rotational states of the upper bound state according to Eq. \ref{eq:fcspc}. Thus, the cross section determined from the slope is the weighted sum of the cross sections of all rovibrational states contributing to the overall emission.

In the conditions of the present experiment, the Xe$_{2}$ excimers are in thermal equilibrium with the atoms of the surrounding gas and many vibrational and rotational states of the molecule are occupied according to the Boltmann distribution. If the collision cross section of the molecule for each of its rovibrational states is $\sigma_{v^{\prime}J^{\prime}},$ the cross section determined from the spectrum broadening is, roughly speaking, a weighted sum of each individual contribution
\begin{equation}
\sigma_{0} \sim \sum_{v^{\prime}J^{\prime}}e^{-\beta E_{v^{\prime}J^{\prime}} }\sigma_{v^{\prime}J^{\prime}}
\label{eq:svjtot}\end{equation} 
The potential energy $V_{u}$ \cite{jonin2002_II} can be very well approximated by a Morse-type potential of the form
\begin{equation}
V_{u}= T_{e}^{\prime} + D_{e}^{\prime}
\left\{ 1- \exp{\left[-\beta_{e^{\prime}}\left(
R-R_{e^{\prime}}
\right)\right]
}
\right\}^{2}
\label{eq:vumorse}\end{equation}
with $T_{e}^{\prime}= 13860\,$cm$^{-1},$ $D_{e}^{\prime}=1717\,$cm$^{-1},$ $R_{e^{\prime}}=3.23\,$\AA, and $\beta_{e^{\prime}}R_{e^{\prime}}=6.734$ \cite{borghesani2007jpb}. Its well strength is such that it can accommodate up to $v^{\prime} =34 $ vibrational levels. At the temperature of the experiment, $T=300\,$K, $k_{\mathrm{B}}T\approx 208.5\,$cm$^{-1},$ the average vibrational quantum number is $\langle v^{\prime}\rangle =3.$ 

The rotational constant of the Xe$_{2}$ in the $(3)0_{u}^{+}$ state is $B_{e}^{\prime}= \hbar^{2}/(2m_{r}R_{e^{\prime}}^{2})\approx 2.47\cdot 10^{-2}\,$cm$^{-1},$ in which $m_{r}\approx 1.09\cdot 10^{-25}\,$kg is the average reduced mass in a sample of natural isotopic abundance \cite{borghesani2007jpb}. It corresponds to a rotational temperature $\Theta_{r}\approx 3.5\cdot 10^{-2}\,$K. For $T=300\,$K, rotational states of high angular momentum are thermally excited. Their population is non-negligible for a rotational quantum number $J^{\prime}$ as high as 250, with average $\langle J^{\prime}\rangle = 80.$ 

Thus, roughly speaking, a number $N_{c}=\langle v^{\prime}\rangle\cdot \langle J^{\prime}\rangle = 240$ of different rovibrational levels, on average, contributes to the measured cross section. If we assume that the excimer size and, hence, its cross section, does not depend too much on its rovibrational state, each of them contributes an amount $\sigma$ to the total cross section $\sigma_{0}$. Hence, $\sigma_{0}\approx N_{c}\sigma .$ In this way a crude estimate for the excimer cross section is easily obtained as $\sigma \approx 8.7 \,$nm$^{2},$ that corresponds to a hard-sphere radius $\rho \approx 16.6\,$\AA\ for the pure gas, and $\sigma\approx 0.88\,$nm$^{2} $ and $\rho\approx 5.3\,$\AA\ for the Ar--Xe mixture.

If one calculates the hard-sphere radius of the Xe$_{2}-$atom interaction as the bond length of the Xe$_{2}$ molecule, $R_{e^{\prime}}$ plus one half of the Xe--Xe (or Ar--Xe) hard-sphere radius, one obtains
$\rho \approx 5.4\,$\AA\ for the pure gas case and $\rho\approx 5.2\,$\AA\  for the mixture case. 

This estimate is in quite good agreement with the value deduced from the analysis of the observed broadening of the spectrum only for the Ar--Xe mixture, whereas it is nearly three times as large for the pure gas. This is an obvious consequence of the value $S=9.84$ of the cross section ratio.

We believe that the difference in the cross sections in the pure gas and in the Ar-Xe mixture is a direct manifestation of the quantum indistinguishability of identical particles.
Let us first consider the pure gas case, in which a Xe atom is assumed to be colliding with the excimer. This binary collision picture is justified by the linear dependence of the spectrum broadening on the gas density.

Just for the sake of visualizing the physical situation, let us consider a slow collision occurring along the direction of the bond axis. For $T=300\,$K, the thermal wavelength of the excimer and of the colliding atom is quite short, of the order of 0.1 \AA, much shorter than the molecular bond length $R_{e^{\prime}}=3.23\,$\AA.  Thus, the wavefunction of the approaching atom is correlated only with that of the closest atom in the molecule whereas it is not correlated over the bond length with the wavefunction of the other atom in the molecule. This predominance of the interaction of the colliding atom with the nearest atom in the molecule during a atom--diatom collision has also been shown in other systems, such as H + H$_{2}$ \cite{krstic1999a}. 

For this reason, the Xe$_{2}-$Xe collision at thermal energy must be rather treated as a Xe--Xe collision. In this case the impact and recoil wavefunctions overlap and interfere and the cross section must reflect the interference pattern of the wavefunctions. This situation bears strong analogies with processes of symmetrical resonant excitation or charge tranfer in which glazing angle collisions are converted to nearly head-on collisions \cite{mb,mason}.

The indistinguishability of identical boson particles leads to the symmetrization of the wavefunctions of the colliding pair. Thus, the scattering amplitude $f(\theta)$ must be replaced by $f(\theta) +f(\theta -\pi)$ and the differential cross section $\mathrm{d}\sigma/\mathrm{d}\Omega =\vert f(\theta) +f(\theta -\pi) \vert^{2}$ is enhanced by a factor 4 with respect to the case of a collision of non-identical particles. The quantum cross section (at low energy) is 4 times larger than the classical one \cite{gasio}. 

Moreover, there are two scattering channels, because the two Xe atoms in the molecule are equivalent. The incoming Xe atom can collide with either Xe atoms of the molecule. Thus, the scattering probability, hence the scattering cross section, is enhanced by a further factor~2.

The situation is completely different in the Ar--Xe mixture because now there is no longer the requirement of indistinguishability of identical particles. In this case, the incoming Ar atom scatters off the molecule that is now to be considered as a single physical entity.
As a conclusion, the scattering cross section for the Xe$_{2}-$Xe collision is 8 times larger than cross section for the the Xe$_{2}-$Ar collision.

In literature, the classical atom--diatom scattering cross section is not uniquely defined in terms of the hard-sphere radii of the participating particles and several formulas can be used. Similarly, the cross section ratio $S$ can be calculated in several different ways. For instance, we quote four different formulas that give comparable results \cite{gislason1990}
\begin{eqnarray}
 S_{1}&= & 8 \left[
 \frac{R_{e^{\prime}}+\rho_{\mathrm{Xe}}}{R_{e^{\prime}}+\left(
 \rho_{\mathrm{Xe}}+\rho_{\mathrm{Ar}}
 \right)/2}
 \right]^{2}
   \label{eq:s1}\\ 
 S_{2}&=  &8\left[
 \frac{4\rho_{\mathrm{Xe}}}{3\rho_{\mathrm{Xe}}+\rho_{\mathrm{Ar}}}
 \right]^{2} \label{eq:s2}\\
S_{3}&= &
8\left[
\frac{6\rho_{\mathrm{Xe}}^{2}}{2\rho_{\mathrm{Xe}}^{2}+\rho_{\mathrm{Ar}}^{2}+3\rho_{\mathrm{Xe}}\rho_{\mathrm{Ar}}}\right]
\label{eq:s3}\\ 
S_{4}& =& 8\left[
\frac{2\rho_{\mathrm{Xe}}}{R_{e^{\prime}}+\rho_{\mathrm{AX}}}
\right]^{2}
\label{eq:s4}
\end{eqnarray}
in which the factor 8 due to quantum effects has been explicitly written.

By inserting the literature values for the hard-sphere radii into the previous equations, the  calculated cross section ratio turns out to be comprised in the range $8.4\leq S\leq 12.$ The experimental value $S_{\mathrm{exp}}=9.84$ is well within this range \cite{mogentale2007}. 

 We note that this picture is very crude as we have not considered, for instance, any asymmetry of the atom--diatom interaction potential that may appear if the collision occurs along a direction forming an angle with the bond axis. However, we believe that asymmetries, if any, should be washed away by the averaging effect of molecular rotations.

In spite of the crudeness of this picture, the agreement of its prediction with the experimental outcome lends credibility to the hypothesis that the different strength of the collision broadening in the pure gas and in the mixture is actually caused by the quantum indistinguishability of identical particles. At the same time, this is a rather surprising result because one is na\"{\i}vely led to expect that such effects manifest themselves at low temperature rather than at room temperature.

 \section{Conclusions}\label{sec:fine}
The present new measurements of the IR emission luminescence of Xe$_{2}$ excimers carried out in an extended density range have confirmed the previous results obtained in a more restricted density interval. 

The correct assignment of the molecular states involved in the bound-free transition and the extension of the investigated density range has allowed to put on firmer theoretical ground the heuristic model developed to explain how cooperative- and multiple scattering effects continuously develop upon increasing the density of the environment of the excimers. In particular, the density dependence of the shift of the luminescence spectrum could be used to obtain an independent measure of the ground-state energy of a quasifree electron in a dense gas. 

Quite surprisingly, the analysis of the differences in the shape of the Xe$_{2}$ spectrum recorded in the pure Xe gas and in a mixture with Ar has put into evidence that quantum effects due to the indistinguishability of identical particles are present even though the temperature of the experiments is fairly high. 

In spite of the success of the model we have developed for interpreting the density dependence shift of the spectrum, several problems are still open. In particular, a better understanding of the broadening of the spectrum is needed to determine reliable values of the optical collision cross section of the excimer. 

Moreover, the future developments of the experiment will pursue the possibility of still increasing the pressure in order to bridge the gap with the liquid. We will also investigate the behavior of the Xe$_{2}$ excimers in mixtures with different RG gases for which the electron--atom interaction is repulsive in order to test the validity of our heuristic model.

\section*{Acknowledgments}We acknowledge the invaluable support of D. Iannuzzi at the  Vrije Universiteit, Amsterdam, The Netherlands.
\bibliography{Xeshift}
\end{document}